\documentclass[aip,apl,reprint,article,floatfix]{revtex4-1}

\usepackage{graphicx}
\usepackage{dcolumn}
\usepackage{bm}

\usepackage{mathtools}
\usepackage{physics}
\usepackage{comment}
\usepackage{xcolor}
\usepackage{amssymb}
\usepackage[normalem]{ulem}
\usepackage[english]{babel}

\newcommand{\Vin}{V_{\rm{ent}}}
\newcommand{\Vout}{V_{\rm{exit}}}
\newcommand{\Ein}{E_{\rm{ent}}}
\newcommand{\Eout}{E_{\rm{exit}}}
\newcommand{\fin}{\varphi_{\rm{ent}}}
\newcommand{\fout}{\varphi_{\rm{exit}}}

\newcommand{\new}[1]{\textcolor{black}{#1}}
\newcommand{\old}[1]{}

\begin{document}
\preprint{APS/123-QED}
\title{Landau Level  Single-Electron Pumping}
\author{E. Pyurbeeva}
\email{eugenia.pyurbeeva@mail.huji.ac.il}
\affiliation{Fritz Haber Center for Molecular Dynamics, Institute of Chemistry, The Hebrew University of Jerusalem, Jerusalem 91904,
Israel}
\affiliation{School of Physical and Chemical Sciences, Queen Mary University of London, Mile End Road, London E1 4NS, UK}
\author{M.D. Blumenthal}
\email{mark.blumenthal@uct.ac.za}
\affiliation{Department of Physics, University of Cape Town, Rondebosch, Cape Town, 7700, South Africa}
\author{J.A. Mol}
\affiliation{School of Physical and Chemical Sciences, Queen Mary University of London, Mile End Road, London E1 4NS, UK}
\author{H. Howe}
\affiliation{University College London, London Centre for Nanotechnology and Department of Electrical Engineering, London, WC1A 0AH, UK}
\author{H. E. Beere}
\affiliation{University of Cambridge, Physics Department, Cavendish Laboratory, Cambridge, CB3 0HE, UK}
\author{T. Mitchell}
\affiliation{University of Cambridge, Physics Department, Cavendish Laboratory, Cambridge, CB3 0HE, UK}
\author{D. A. Ritchie}
\affiliation{University of Cambridge, Physics Department, Cavendish Laboratory, Cambridge, CB3 0HE, UK}
\author{M. Pepper}
\affiliation{University College London, London Centre for Nanotechnology and Department of Electrical Engineering, London, WC1A 0AH, UK}

\date{\today}

\begin{abstract}
    We present  \old{the first} \new{a} detailed study of the effect of a strong \new{perpendicular} magnetic field on single-electron pumping in a device utilising a finger-gate split-gate configuration\new{, characterised by high pumping precision}. In the quantum Hall regime, we demonstrate electron pumping from Landau levels in the leads, where the measurements exhibit pronounced \old{oscillations} \new{fluctuations} in the lengths of the pumping plateaus with magnetic field,  reminiscent of  Shubnikov-de Haas oscillations. \new{In order to explain this similarity, we introduce a new physical model for the dynamics of the operation of a single-electron pump. We show} \old{This similarity indicates} that the pumping process is dependent on the density of states of the 2D electron gas \new{in the leads} \old{over a} \new{within a single}  narrow energy window. 
    \old{Based on these observations, we develop a new theoretical description of the operation of single-electron pumps which for the first time} \new{Applying our model to experimental data} allows for the determination of  physical parameters of the pump and its operation; such as the capture energy of the electrons, the broadening of the  quantised Landau levels in the leads, and the quantum lifetime of the electrons. 

\end{abstract}
\maketitle
\section{Introduction}
Technological applications, such as quantum information processing, nanoelectronics, and electron quantum optics require the control of individual electrons with a high degree of accuracy\cite{RevModPhys.85.1421,zim,Bauerle2018,Kaneko_2016,PhysRevLett.110.136802,articleDiVin,PhysRevB.80.113303,6491ca12744642e6aac757534e34609a,Giblin_2019,Yamamoto2012,Ji2003,Ubbelohde2015,Gumbs2009,Fletcher_2013,Kaestner2015, Ubbelohde2023,Brange2023}. AlGaAs high frequency single-electron pumps\cite{Blumenthal2007, Kaestner2008} \new{as well as silicon pumps\cite{sil1,sil2,sil3}} capable of delivering quantised charges, have been extensively studied for this purpose in the past decades, with considerable advancements made\cite{Giblin_2023,Blumenthal2023}. In silicon based electron\cite{articleyam} and hole pumps\cite{articlegen}, there has been growing interest in single-electron pumps that utilise localised states, such as dopants implanted in a channel, instead of electrically defined islands\cite{Moraru2007,roche2013two,2012NanoL..12..763L}. Notably, the high-speed operation of a single donor pump has been demonstrated\cite{tettamanzi2014charge}, highlighting the potential of this approach. 

In the characterisation of deterministic single-electron pumps, the Universal Decay Cascade (UDC) model presented in the seminal work by \emph{Kashcheyevs et al}\cite{Kashcheyevs2010} remains the only paradigm. \new{Its main result is the equation\cite{Kashcheyevs2010, Howe2021}:
\begin{equation}
\label{eq-2ex}
    I=ef \sum_n \exp \left[ -\exp \left(-\alpha_n(\Vout -\delta_n) \right)  \right]
\end{equation}
which} \old{It} gives good agreement with experimental data and allows for the determination of a set of parameters \new{$\alpha_n, \delta_n$ (where $n$ is the plateau number), acting as a ``fingerprint''} \old{(a ``fingerprint'') that quantifies} \new{quantifying} the accuracy and dynamics \new{of an individual device}. The model, however, is limited in its ability to explain experimentally observed physical phenomena. This limitation needs to be overcome as more complex devices are utilised as electron pumps\cite{Rossi2018}.

We present experimental observation of a split-gate finger-gate (SFG) electron pump\cite{Howe2021} operating in a perpendicular magnetic field. The evolution of the pumped current is studied as the magnetic field is swept from $1$\,T to $9$\,T. \new{Some studies of the effect of a magnetic field on electron pumping have suggested monotonic lengthening of the pumping plateaus, and thus increased pumping precision\cite{Wright2008, Kaestner2009, Leicht2010, HANIEF2024e02150}, while others\cite{Leicht2011, Kataoka2011} observed oscillations of the plateau lengths. \emph{Leicht et al.}\cite{Leicht2011} compared these oscillations to the Hall resistance of the pump as a function of magnetic field. Our observations, making use of the high precision of the SFG pump configuration\cite{Howe2021}, are consistent with these results\cite{Leicht2011} and show that the plateau length evolution} \old{We find that the pumping plateaus do not lengthen monotonically with field, as has been previously suggested\cite{Wright2008, Kaestner2009, Leicht2010, HANIEF2024e02150}, but instead their evolution} is closely correlated with the Shubnikov–de Haas (SdH) oscillations \new{we measure} in the same device. This allows us to introduce a new physical model \new{for the operation of the pump}, \new{which is in agreement with the results of} the UDC model, \new{and separates} the individual properties of the device from the effect of the magnetic field. It gives new insight into the dynamics of the pumping process, describes the data qualitatively and allows us to determine certain numerical parameters of the system, such as the capture energy of the pump and the electrons' quantum lifetime. 

\section{Experimental Data}
\begin{figure*}
\includegraphics[width=\textwidth]{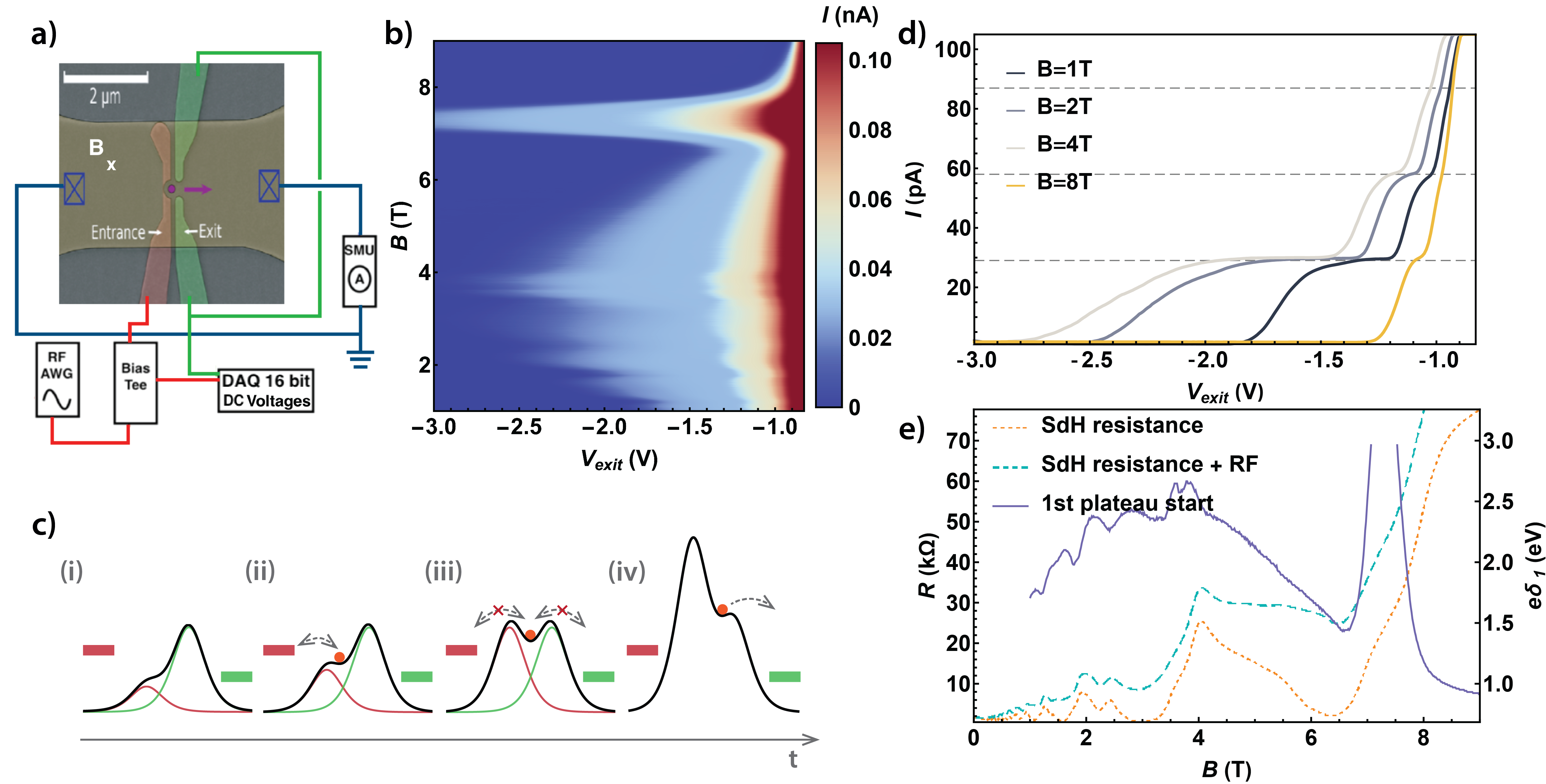}
\caption{a) A false colour SEM image of the SFG device with experimental set-up. \new{Two terminals are shown, however the device contains a four terminal Hall geometry allowing for routine measurement of the quantum Hall effect and SdH.} b) 2D colour map showing the dependence of the pumped current on the magnetic field \new{$B$} and exit voltage $\Vout$. c) A set of schematics illustrating a single pump cycle in a SFG pump. (i) At low momentary values of $e \Vin (t)$ the potential profile presents a single maximum, preventing direct current from the source to the drain. (ii) As $e \Vin (t)$ increases, \old{a} confined state\new{s} form\old{s} in the potential profile\new{, and exchange electrons with the source}. (iii) Further increase of $e \Vin(t)$ isolates the confined electron\new{(s)} from both the source and the drain, \new{such that no further electron can enter or decay from the QD}. (iv) \old{The} \new{All} confined state\new{s are} \old{is} destroyed and electron\new{s are} \old{is} deposited into the drain. d) Pumping traces in $\Vout$ for several values of $B$. The lengthening of the plateaus can be observed initially, in agreement with previous work \cite{Howe2021}, but it is not monotonic in $B$. e) The starting voltage of the first plateau, \new{$e$}$\delta_1$ (solid purple) vs. the Shubnikov-de Haas effect in the same sample with RF excitation on the finger-gate (dashed turquoise) and without (dashed orange). }
\label{fig1}
\end{figure*}

The SEM image of the pump and the schematic of the experiment are shown in Fig.\ref{fig1}a, with further details \new{on fabrication and measurement} given in the Methods section. \new{The pump in our experiment was run at constant RF amplitude ($V_{\text{Amp}}=300$\,mV), frequency ($f_{\text{RF}}=180$\,MHz), entrance gate DC offset voltage ($\Vin=-600$\,mV) and source-drain bias ($V_{\text{SD}}=100$\,mV). The magnetic field $B$ and the exit gate voltage $\Vout$ were set as variable parameters. Fig.\ref{fig1}b shows a pumpmap\cite{Wright2010} of the current as a function of $\Vout$ and $B$}.  

The principle of device operation is illustrated in Fig.\ref{fig1}c \new{(i) - (iv), which shows the time evolution of the} cross-section of the spatial 3D potential created by the gates \new{(we show $e\varphi$, the negative of the electrostatic potential, illustrating the potential energy of electrons)}. At the beginning of the cycle (i), the left entrance barrier is low and the dot potential profile has a single maximum. As the entrance barrier potential increases (ii), a local minimum appears in the potential, creating confined states which can \new{exchange and become}\old{be} occupied \old{by}\new{with} electrons from the source (capture stage). With further increase \new{of left barrier height} (iii), the confined states become isolated from the source and move with the evolution of the potential profile. \new{Here, no further electrons can enter the dot and the decay of electrons back to the source is suppressed.} Finally (iv), the confined states are destroyed and \new{all} electrons are ejected into the drain.

\new{We note that Fig.\ref{fig1}c merely illustrates the configuration of the electrostatic potential of the pump.} The number of pumped electrons per cycle at constant RF frequency and amplitude \new{depends on}\old{by}  both the DC gate voltages\new{, as well as the magnetic field (Fig.\ref{fig1}b). In our case, the bias voltage\cite{Howe2021,Blumenthal2023} is such that we assume the ejection to always be complete, therefore the pumping depends on the density of states in the source only. Exploring this relation is the primary focus of the paper.} 

Fig.\ref{fig1}d shows four pumping traces at different magnetic fields for a fixed entrance gate offset voltage and RF parameters. The pumped current evolves as a function of the exit gate voltage $\Vout$, with the integer number of pumped electrons increasing as the DC voltage on the exit barrier is made more positive (the barrier is lowered) -- this is a common feature of electron pumping\cite{Blumenthal2007,Kaestner2008}. 

\new{As a function of magnetic field, the first three traces in Fig.\ref{fig1}d show a trend}\old{Fig.\ref{fig1}b shows a pumpmap of the current as a function of $\Vout$ and magnetic field $B$\cite{Wright2010} with the entrance gate DC voltage fixed at $V_{\text{ent}}=-600$\,mV, the source-drain bias $V_{\text{SD}}=100$\,mV, and the RF amplitude and frequency $V_{\text{Amp}}=300$\,mV and $f_{\text{RF}}=180$\,MHz. On observation, the plateau lengths depend strongly on the magnetic field, and the general trend over the field $1-3$\,T is} towards plateau lengthening, in agreement with  previous studies\cite{Wright2008, Kaestner2009}. The plateau length dependence is however non-monotonic, \new{as can be seen from the $8$\,T trace}, but instead oscillatory. \new{This behaviour is further emphasised in Fig.\ref{fig1}b, a full magnetic field dependent pump map, where a dramatic resonance peak at around $7.2$\,T is visible.} \old{including a dramatic resonance peak at around $7.2$\,T.} 

These oscillations are reminiscent of magnetic field dependencies seen in the Shubnikov-de Haas effect\new{, similarly to the Hall resistance in \emph{Leight et al.} \cite{Leicht2011}} \new{To make a comparison,} we study the SdH effect in the pump by measuring the longitudinal resistance of the device as a function of magnetic field, first with the finger gate grounded (Fig.\ref{fig1}e, orange) and then with an RF signal on the finger gate (Fig.\ref{fig1}e, turquoise) with the parameters of the RF signal set as used during pumping. 

\new{The purple plot in Fig.\ref{fig1}e shows the potential energy barrier height (absolute value of the exit gate voltage)} \old{These SdH plots in the quantum Hall regime are compared to the exit gate voltage} corresponding to the start of the first pumping plateau (current is at $1/e$\new{, where $e$ is the natural exponent,} of the expected plateau value, see Eq.\ref{eq-2ex}) as a function of magnetic field. \old{(Fig.\ref{fig1}e, \old{blue}, \new{purple})} \new{While the three plots in Fig.\ref{fig1}e are far from identical, there is significant similarity in their peak structures that requires explanation.} 

\old{The SdH plot without the RF signal differs from assessment measurements in the same wafer with a standard Hall bar. This difference can be attributed to the geometry of the mesa in the pump deviating substantially from a standard Hall bar geometry. The inclusion of Ti/Au gates as well as the fabrication process itself also affect the wafer, altering the electron density and mobility. The application of the RF signal further influences the magnetic field dependence of the device resistance, however both retain the general peak structure of the signature SdH.}

As the magnetic field oscillations in the SdH effect originate in the Landau quantisation of the electron states, we make the conjecture that similarly, the pumping of electrons depends on the population of the density of states in the source of the pump. \new{The difference between the pure SdH signal (Fig.\ref{fig1}, orange) and the first plateau starting potential (Fig.\ref{fig1}, purple) can be explained by the presence of RF signal (Fig.\ref{fig1}, turquoise), as well as the exact mechanism of pumping. The change in electron density observed when comparing the orange and turquoise plots is attributed to the voltage applied by the RF signal generator to the finger gate. This voltage locally modifies the electron density, either increasing or decreasing it depending on the specific characteristics of the applied signal. In this experiment, we replicate the RF signal used during the pump cycle, which generates an overall potential profile averaged over the full RF cycle. This profile results in a depletion of electrons. There is also present an expected increase in the longitudinal resistance. This is due to the reflection of one or more edge states by the RF gate\cite{PhysRevB.48.8840}.} \old{The work expands on this assumption} 

\new{The model and simulation to follow expands on this experimental observation}.
\section{The model}

\subsection{Motivations \old{and criteria for the model}}
A single-electron pump is a complex open quantum system \new{incorporating both particle and energy exchange, a time-dependent Hamiltonian, multiple competing time-scales, out-of-equilibrium conditions and multi-electron processes, all of which make it a serious challenge for theoretical description\cite{Pyurbeeva2022a}.}\old{ with many aspects which make it a serious challenge for theoretical description\cite{Pyurbeeva2022a}: both particle and energy exchange being present, a time-dependent Hamiltonian, multiple competing time-scales, out-of-equilibrium conditions and multi-electron processes.} Combinations of these \new{effects} are 
\new{studied in topics such as far-from-equilibrium thermodynamics in nanodevices\cite{Pyurbeeva2023}, as well as in applications of driven quantum dots such as quantum heat engines and refrigerators\cite{Juergens2013, Hino2021, Monsel2022, Monsel2023}.}

\old{studied in topics such as driven quantum dots, for applications such quantum heat engines or refrigerators\cite{Juergens2013, Hino2021, Monsel2022, Monsel2023}, or far-from-equilibrium thermodynamics in nanodevices\cite{Pyurbeeva2023}. }

\old{However, single-electron}\new{Single-electron} pumps \new{also} exhibit additional complexity due to qualitative changes, such as the creation and destruction of confined states, as well as  strong system-bath coupling associated with it. \old{alongside quantitative changes, like cycling of addition energy.} \new{Quantitative changes, such as the cycling of addition energy further complicate the system.} \old{Recent research\cite{Schulenborg2023} has focused on dynamical analysis from an open quantum system perspective, including a study on two-electron emission processes from a quantum dot, however such approaches are computationally involved. Our investigation, focusing on the capture side of the cycle and incorporating a magnetic field, takes a more qualitative and intuitive approach.}

\new{The} \old{The main result of the UDC model is a} double-exponential fitting equation for the pumping curves\cite{Howe2021} \new{of the UDC model (Eq.\ref{eq-2ex}) is the primary theoretical approach to the description of single-electron pumps.} \new{It} gives a good fit to the \old{experimental data}\new{pumped current} in a wide variety of electron pump geometries\cite{Howe2021} \new{and allows to compare devices operating in identical experimental conditions}. \old{but} \new{Despite} later expansions\cite{Kashcheyevs2012, Kashcheyevs2014} on the model, it does not\new{, however,} link \new{the ``fingerprint'' coefficients $\alpha_n, \delta_n$} explicitly to any key physical parameters such as temperature, magnetic field, RF signal amplitude, \new{or} frequency. \new{Such a connection can be found from the explicit consideration of a physical model. Recent work\cite{Schulenborg2023} performed a dynamical analysis for a driven quantum dot from an open quantum system perspective, including a study on two-electron emission processes, however such approaches are computationally involved and highly model-dependent. }

\new{We propose a \old{more} qualitative and intuitive approach, using experiment as a starting point.} Our experimental data (Fig.\ref{fig1}b) provides \old{the first} \new{a} systematic high-precision study of the dependence of \old{the}\new{quantised} pumping on \old{an external parameter,} the magnetic field. \new{Our observations serve as a basis for developing a physical model of the dynamics of the electron pump}.

\subsection{\new{Observations}}

\begin{figure}
    \includegraphics[width=\linewidth]{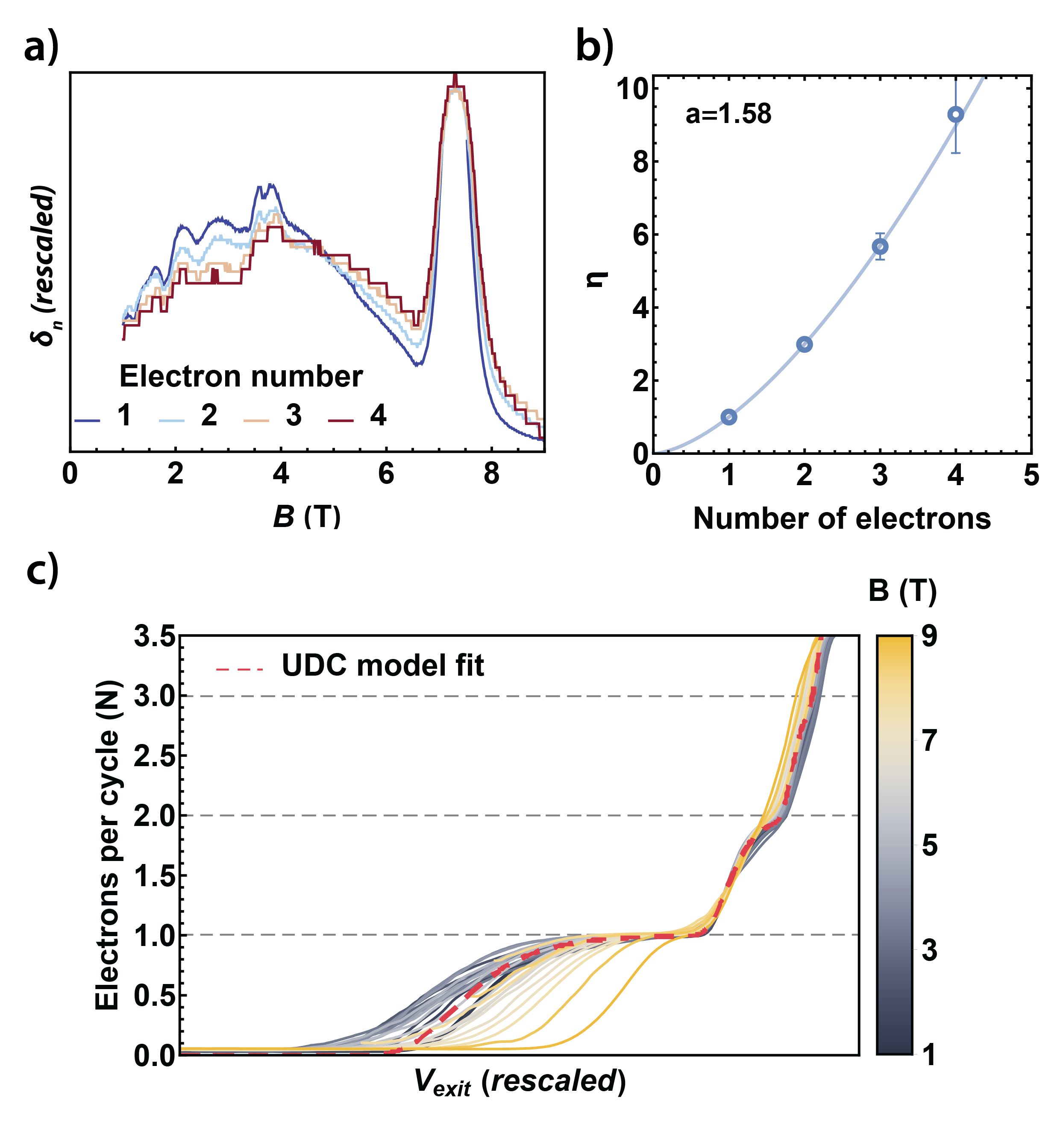}
    \caption{a) The \old{onset of pumping at 1/e of the true integer plateau of} \new{pumping onset voltage $\delta_n$ for} the first four plateaus ($n=1...4$) linearly scaled to coincide with each other. b) The calculated scaling factors of $\delta_1$--$\delta_4$. c) \old{Normalised}  \old{p}\new{P}umped current vs. $\Vout$ for varying magnetic fields from $B=1$\,T to $B=9$\,T in $0.5$\,T steps\new{, horizontally rescaled by the magnetic field-dependent parameter $\lambda(B)$}. Red dashed line shows the UDC fit \new{to the averaged trace}. }
    \label{fig2}
\end{figure}

\old{We now construct a model to describe our findings.}
\new{Before constructing a model, we further detail our experimental observations. Fig.\ref{fig1}e compares the onset of pumping of the first plateau as a function of $\Vout$ \new{(dark blue)}, which corresponds to the current reaching $1/e$ of the true expected integer plateau\new{, where $e$ is the natural exponent}, to the Shubnikov-de Haas resistance in B -- turquoise (with RF) and orange (No RF). The pump map in Fig.\ref{fig1}b shows the same peak pattern for all plateaus (n=1,2,3,4).}

\new{We proceed with a direct comparison of the first four plateaus.} In Fig.\ref{fig2}a \new{we plot} the \new{onset} voltages \new{for the first four pumping plateaus, $\delta_1$--$\delta_4$ (voltage points} \old{(}  \new{from} $n=1$ (blue) to $n=4$ (red) \old{are given} as a function of $B$\new{,} \old{. These line scans were taken from Fig.\ref{fig1}b and} linearly scaled \new{to minimise the sum of the squares of differences between the corresponding data points from each pump plateau for all B.} \old{by a factor of ($\eta$), shown in Fig.\ref{fig2}b.} The scaling factors \new{$\eta(n)$} follow a power law in $\eta(n)\approx n^a$ with $a=1.58$.

$\delta_1$--$\delta_4$ coincide (overlap of all curves in Fig.\ref{fig2}a) up to a linear transformation \new{indicating that all pumped plateaus follow the same oscillatory pattern reminiscent of the Shubnikov-de Haas oscillations, instead of a separate magnetic field dependence for each $\delta_n$. The pump plateaus all dilate in the same way, either contracting or expanding  in $\Vout$ as the magnetic field is varied.}

\new{This allows us to introduce} a single parameter \old{$\delta(B)$}  \new{$\lambda(B)$} governing the overall length of the pumping trace, \new{such that the UDC equation (Eq.\ref{eq-2ex}) can be written as:
\begin{equation}
    I=ef \sum_n \exp \left[ -\exp \left(-\alpha_n(\Vout -\lambda(B) \delta'_n) \right)  \right]
\end{equation}
where the coefficients $\delta'_n$ are magnetic field-independent.} \old{rather than the start of each plateau having separate magnetic field dependencies.} \\
\new{To further substantiate our claim that the entire pump map is represented as a single pumping trace dilating and contracting with magnetic field as a whole, we present Fig.\ref{fig2}c, which} shows \old{the} pumping traces at different magnetic fields \old{stretched} \new{dilated} by \old{the length factor} \new{$\lambda(B)$.} \new{Additionally, Fig.\ref{fig2}c shows the UDC fit (dashed red) to the average of the coinciding pumping traces, which comprises the} pumping trace inherent to the device \new{at $B=0$ with $\lambda(0)=1$ and the experimental conditions,,such as frequency and temperature, remaining constant}. \new{The pump traces shown here are taken from a pump map operating in the UDC regime. For a full $V_{\text{ent}}$ vs $V_{\text{ext}}$ pump map of this device please refer to the following papers\cite{Howe2021,Blumenthal2023} where the same device was measured.} \old{further supporting this claim, and showing a ``true''.} 

\new{From these experimental observations, we can thus conclude that the magnetic field dependence of pumping can be described as that of a single device-dependent pumping trace, dilating and contracting with magnetic field by the length factor $\lambda(B)$}. \new{This dilation strongly resembles the SdH oscillations of longitudinal resistance of the 2DEG}, which is  proportional to the density of states at the Fermi-level \old{as discussed previously} (Fig.\ref{fig1}e). \old{We can conclude that the pumping process not only depends on the density of states in the source electrode, but on the density of states  at a single energy -- if $\delta(B)$ was governed by electron exchange processes at different energies, it would be a convolution of $n(B, E)$ at different values of $E$, and the contributions from different Landau levels would be washed out. The sharp Landau level signatures observed indicate that the entire pumping effect is determined by exchange processes confined to a very narrow energy window.} 

\subsection{The 0-DIP model and dot parameters}
\begin{figure}[h!]
    \centering
    \includegraphics[width=\linewidth]{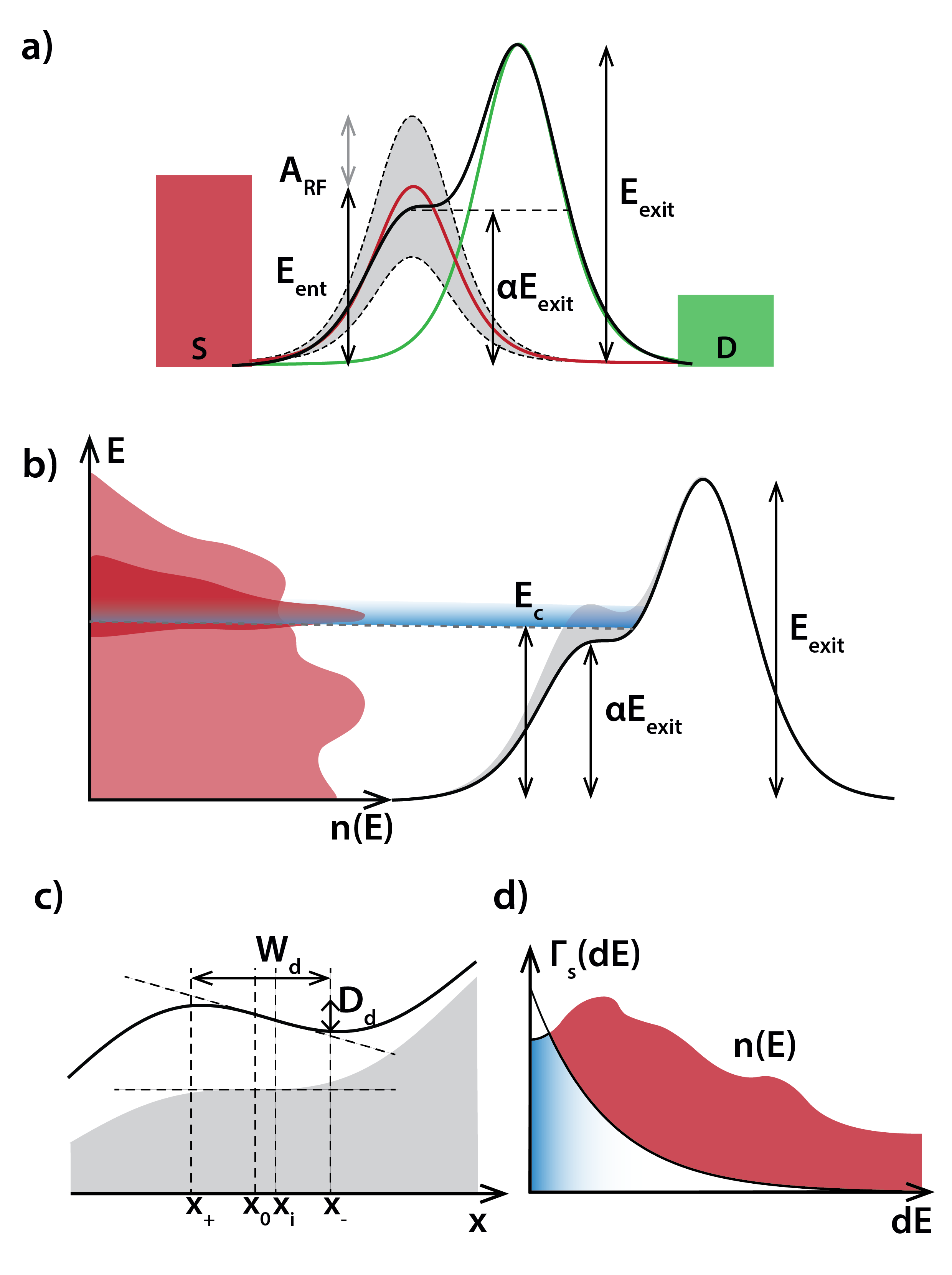}
    \caption{a) A labelled schematic of the \old{theoretical model} \new{potential profile parameters}. \new{The green and red peaks show the exit potential profile and the constant offset of the entrance potential respectively. The black curve is the 0-DIP potential, realised at one of the momentary values of $\Ein$, which lie within the grey shaded area with the width being the amplitude of the RF signal $A_{RF}$.} b) The device potential profile \old{in} \new{in the vicinity of} the 0-DIP configuration. \old{T} \new{As $\Ein$ goes past the the 0-DIP configuration (t}he grey shaded profile\new{),} \old{shows the quantum dot at the cut-off point where it becomes isolated from the source electrode.} \new{the first confined state appears at $E_c$, after which the coupling with the source falls exponentially. All electron exchange processes occur within the blue shaded energy range.} c) Schematic showing the zoomed-in region of the \old{minima in the} QD potential \new{close to the 0-DIP point} with the length scale dimensions ($W_d$ width, and $D_d$ depth) indicated. The grey area indicates the 0-DIP configuration. \new{$x_+$ and $x_-$ are the local maximum and minimum respectively, $x_i$ is the inflection point, $x_0$ is the inflection point in the 0-DIP configuration.  } d) The schematic showing the change in tunnel rate $\Gamma$ and \new{an arbirtary} source electron density $n(E)$ as the energy of the entrance barrier is changed ($\dd E$). \old{e) A schematic of the effect of $\Eout$ on the pumping. Lowering the energy of the exit barrier leads to an increase in the size of the dot at the 0-DIP potential configuration, which lowers energy of capture on $\Vin$ given by $\alpha\Delta\Eout$. In the case of a very narrow band in $n(E)$ (shown in dark grey) the change in capture energy could lead to a reduction of the pumping current despite an increase in the size of the dot.} }    
    \label{fig3}
\end{figure}
We now propose a  physical model \old{based on} \new{explaining} our \new{experimental} observations.

\new{The main physical premise of the UDC model\cite{Kashcheyevs2010} is that the quantum dot is ``loaded'' with electrons while its energy levels are beneath the Fermi level, and once the quantum dot is energetically raised above it, non-equilibrium relaxation, or backtunnelling, occurs until some number of electrons remain to be pumped into the drain. However, this is not corroborated by our experiment.}

\new{Electron pumping is inherently governed by electron exchange between a 2DEG and a quantum dot. Such processes are commonly dependent on the density of states in the electrode\cite{Beenakker1991}, which, in our case, can be seen in the SdH resistance. However, the sharp Landau level signatures in $\lambda(B)$, the overall trace length factor, can only be observed if all electron exchange processes determining the pumping at a given magnetic field are confined to an energy window much narrower than the width of the Landau level, in contradiction to the usual description of the UDC model.}

\new{To explore the physical nature of electron pumping and explain the origin of this energy confinement of electron exchange, we study} \old{We utilise} a 1D \new{model of} the electron \new{potential} energy profile. \new{When} voltages $\Vin$ and $\Vout$ \new{are applied} to the entrance and exit electrodes, \new{the protential energy profile $E(x)$ is a sum of two} \old{as} smooth differentiable peak functions $ \Ein \varphi_{\rm{ent}}(x)$ and $ \Eout \varphi_{\rm{exit}}(x)$ with continuous first and second derivatives. \new{The peak maxima are} \old{The maximal values of the electron energies are} \new{given by} $\Ein$ and $\Eout$, which depend linearly on the voltages applied (Fig.\ref{fig3}a) \new{ and $\varphi_{\rm{exit}}(x)$ and $\varphi_{\rm{ent}}(x)$ are the dimensionless spatial potential profile functions with maximum values of 1}.
If the peaks are sufficiently close to one another, for every value of $\Eout$ there exist \new{a value} \old{two values} of $\Ein$ for which the total potential profile exhibits an inflection point with a zero derivative \old{with the inflection points} on the source side \old{for low  $\Ein$ and drain side for high $\Ein$} \new{of the potential profile}. We call \old{the former} \new{this} configuration \new{(Fig.\ref{fig3}b)} the \emph{0-DIP potential} \old{(Fig.\ref{fig3}b)} \new{(for ``zero derivative inflection point'')}. The value of \old{$\Ein$ at} \new{the inflection point energy at the} the 0-DIP potential \new{configuration} is proportional to $\Eout$ \new{(and therefore $\Vout$)}, where we define the proportionality coefficient as $\alpha$.\\
The significance of the 0-DIP potential is that it approximately separates two configurations of the dynamically defined dot -- that with no confined electron states\old{($\Ein<\alpha \Eout$)}, and that \old{other} containing one or multiple confined states. We assume that \new{the first} confined state\old{s form} \new{forms} soon after $\Ein$ passes the 0-DIP potential \new{at an energy $E_c$, which is close to $\alpha \Eout$}. A crucial point in our model is that a further increase of $\Ein$ \old{ with the application of the RF signal} \new{during the RF cycle} leads to a rapid reduction in the tunnel coupling between the now-formed \new{confined states} \old{dot} and the source (Fig.\ref{fig3}c)\new{, and} after a small increment $\dd E$ in entry barrier height, the dot is \new{effectively} completely isolated, not only from the drain, but also from the source (grey shaded profile in Fig.\ref{fig3}b). \new{This means that $E_c$, the energy of the formation of the first confined state can be thought of as the capture energy, since all electron exchange processes occur within a narrow energy window after this energy as indicated in Fig.\ref{fig3}b) by the blue shaded strip.}      
\old{The value of $\dd E$ is assumed small in comparison to the characteristic broadened width $\Gamma$ of the Landau levels. This assumption contributes to the explanation of the oscillatory behaviour of the pumped current vs magnetic field in Fig.\ref{fig1}b  and is a requirement of our model. }\\
To \new{substantiate this claim,} \old{demonstrate the validity of this 0-DIP model, we investigate the evolution of the geometry of the quantum dot during the pump cycle. W} \new{w}e define two fundamental dimensions of the dot: \new{$W_d$ --} the width of the entrance potential barrier \new{and the characteristic size of the dot (the distance between the local minimum and maximum in the potential profile)}, \old{$W_d$} and \new{$D_d$ -- } the depth of the dot \old{$D_d$} \new{and the height of the potential barrier (the energy difference between the local minimum and maximum in the potential profile)} -- see Fig.\ref{fig3}c -- and \new{study their evolution} \old{model the change in these dimensions} as a function of $\dd E$ in a one-dimensional system \new{close to the 0-DIP point}. Analytically making use of Taylor expansion (see Methods) we find the following two expressions for the dot parameters: 
\begin{equation}
\label{pars}
    \begin{dcases}
    W_d=C_W \sqrt{\frac{\dd E}{\Eout}}\\[5pt]
    D_d=C_D \sqrt{\frac{(\dd E)^3}{\Eout}}
    \end{dcases}
\end{equation}
While the exact form of the dependence will be different for a two-dimensional case and will depend on the geometry of the electrodes, qualitatively it will follow the 1D case. \old{, with the barrier height and width increasing with $\dd E$ and decreasing with $\Eout$. The effect of dot size on pumping can be seen in the well-known universal feature of the pumping traces -- less negative $\Vout$ (lower $\Eout$) leading to more electrons being pumped per cycle, which intuitively agrees with a geometrically larger quantum dot. Additionally, t} \new{T}he increase of $W_d$ with $\dd E$ supports the \old{ initial assumption} \new{statement} of the dot only being coupled to the source in a narrow energy window, as the coupling to the source $\Gamma_S$ depends exponentially on $ W_d$ (Fig.\ref{fig3}d). \new{Additionally, an increase of $ W_d$ with lower $\Eout$ agrees with more electrons per cycle being pumped with less negative $\Vout$.} 

\new{We note that the different physical model is not in contradiction with the results of the UDC model, as Eq.\ref{eq-2ex} does not change with the range of energies at which backtunnelling occurs. Furthermore, we are not attempting to describe the exact dynamics of electrons around the 0-DIP point, between the formation of confined states and the quantum dot being decoupled from the source -- such work can form a separate study. We simply claim that however complex the exchange dynamics, they are limited to a small energy window, and can be used to probe the density of states in the source, as the capture probability depends on it.}

\new{Figure \ref{fig3}b shows the energy profile of the pump close to the 0-DIP point and a source with an arbitrary density of states. A change in $\Eout$ leads to two effects -- a change in the inflection point, $\alpha \Eout$, and therefore capture energy $E_c$, and a change in the size of the dot, and therefore the probability of capture. Since the pumping trace length is a function of magnetic field $B$ only, we can conclude that the first effect is not significant and the change of capture energy over our line scan is small compared to the width of the Landau level (if it were not the case, we could observe a decrease of the number of pumped electrons with an increase of exit voltage, see Fig.\ref{fig3}b, dark red). This means that the capture energy is effectively constant throughout the experiment, or doesn't change in comparison to the width of the Landau level.}

\new{An important distinction is that while in our model pumping depends on the density of states at the capture energy, which is constant and a property of the pump, the SdH resistance gives the density of states at the Fermi energy, which is a function of magnetic field. This will be the source of some misalignment between the data in Fig.\ref{fig1}e.}

\section{Data analysis -- density of states and capture energy simulation}

\old{We now apply the developed 0-DIP our physical model to the experimental data. We simulate the change in  how the capture energy.} \new{We now simulate the magnetic field  dependency of the pump cycle, as shown in Fig.\ref{fig1}b, by investigating how the capture of electrons at  $E_c$ as defined in the 0-DIP model is impacted by the change in the density of states (Landau levels) and Fermi energy in the leads as the B-field is swept from $1$\,T to $10$\,T. Our working assumption is that in order for the pump to deliver electrons from the source to drain, the capture energy $E_c$, must be in resonance with a Landau level in the source that is indeed populated with electrons.} \new{The simulation will also allow} us to determine certain numerical properties of the pumping mechanism in high fields and validate our \new{assumptions}.

\begin{figure*}
    \centering
    \includegraphics[width=\textwidth]{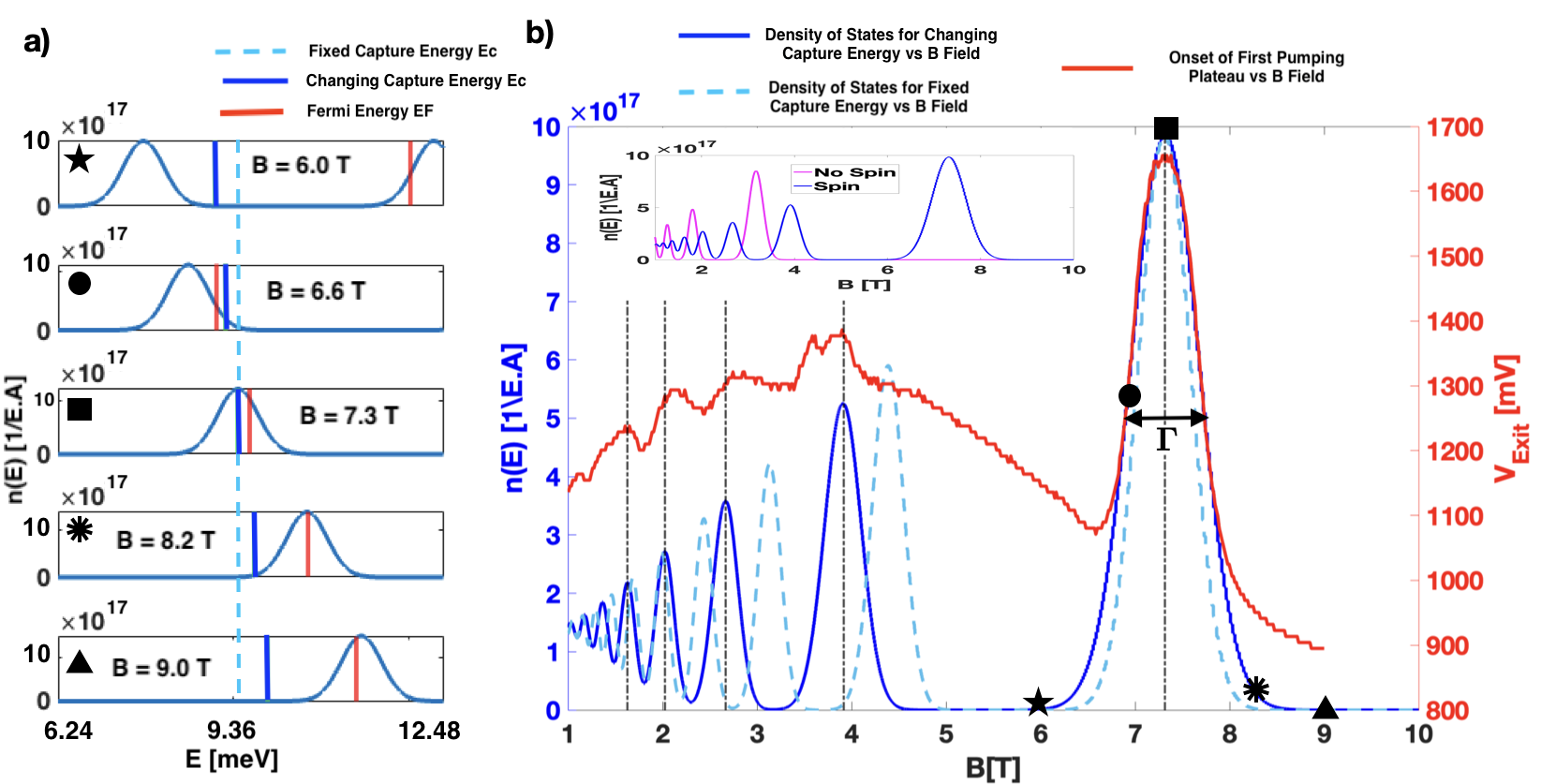}
    \caption{a) Schematics of the evolution of the spin-split first Landau level for $B$ values chosen around the main resonance peak centred at $B=7.3$T. The black shape markers indicate the chosen $B$ values on the resonance peak in panel b. The Fermi energy is indicated by the red vertical lines, the turquoise dotted line indicates a fixed capture energy and the blue line a changing capture energy, with $B$.  b) Onset of 1st plateau pumping (experimental data) in red taken directly from Fig. \ref{fig1}e. Blue and turquoise plots indicate the simulated density of states for changing capture energy and fixed capture energy respectively vs $B$ field. $\Gamma$ shows from where in the experimental data (red) plot the Landau broadening was determined. Inset: density of states for changing capture energy vs $B$ field, shown considering electron spin splitting (blue) and ignoring spin splitting (pink).}
    \label{figLand}
\end{figure*}

\old{The electron pump, defined in a quasi one-dimensional  restriction etched into a 2DEG AlGaAs mesa with multiple Ohmic contacts, allowed for routine Quantum Hall and SdH measurements to determine the electron density of the pump mesa.}\new{First we explicitly determined the electron density of our pump using routine Quantum Hall and SdH measurements.}  Initially, all gates were grounded, and then measurements were repeated with the finger gate oscillating at the same frequency and amplitude used during the pump's operation. These measurements revealed an electron density of $n_D = 1.53\times 10^{15}$\,m$^{-2}$, which corresponds favourably with what was measured by the growers ($n_D=1.87\times10^{15}$\,m$^{-2}$) in the Cavendish Laboratory on the stock wafer before any fabrication on the wafer.

From the SdH oscillations and measured electron density we were in a position to determine the filling factors $\nu$. We observe that the final minimum in the SdH data (Fig.\ref{fig1}e orange), where the Fermi level is in a localised state, corresponds to all electrons in the first Landau level ($\nu=2$, including spin). This aligns with the minimum in the pumping data at $6.7$\,T, occurring just before the onset of the significant resonance peak at $7.3$\,T (Fig.\ref{fig1}e). The last minimum in the SdH data shows a stronger coincidence with the minima in the pumped data if the RF gate when measuring the SdH was switched on. \\
\old{Using}We \new{could} then \old{simulated} \new{simulate} using code written in Matlab, the movement of the Landau levels and the Fermi energy governed by the determined density of electrons \old{above} as the magnetic field was changed. We focus on the region of the resonance peak at B values of $6.0$\,T, $6.6$\,T, $7.3$\,T, $8.2$\,T and $9.0$\,T. 

First we determined the Landau level broadening (fwhm), a requirement for observing the quantum Hall effect\cite{Potts_1996}, which is given by $\Gamma=\frac{\hbar}{\tau_i}$, where $\tau_i$ is the quantum lifetime, \new{the mean time a carrier remains in a momentum eigenstate before scattering\cite{PhysRevB.96.035309}}. $\Gamma$ is determined from \new{the width of} the resonance peak (red line, Fig.\ref{figLand}b), which as previously discussed in the 0-DIP model, corresponds to the dot probing the Landau level with filling factor $\leq 2$ (including spin), \new{as it moves past the capture energy point with $B$} (Fig.\ref{figLand}a). We obtained \old{$\Gamma=1.35\times10^{-22}$\,J} \new{$\Gamma=0.84$\,meV}, corresponding to a lifetime of $\tau_i=0.78$\,ps. \old{This lifetime, the mean time a carrier remains in a momentum eigenstate before scattering\cite{PhysRevB.96.035309}, is an order of magnitude shorter than typical time scales from standard quantum Hall measurements\cite{MARTIN1988323}.} \new{Our determined $\tau_i$ corresponds favourably with what is quoted in literature\cite{quantum1,quantum2,quantum3}.} In the pumping regime, introducing the RF signal disrupts the system, increasing scattering and shortening the quantum lifetime. Further investigations will explore the RF frequency and amplitude's impact on $\tau_i$. 

The movement of the Landau levels (\old{black}\new{blue} peaks) and Fermi level (red lines) are shown in the selection of plots in Fig.\ref{figLand}a with the width of the Landau peaks as determined above. At the resonance peak at $7.3$\,T the Fermi level despite not being centered in the lowest Landau level is contained within it\new{, indicating all electrons are indeed populating the first Landau level.}

With this finding we then determined the capture energy of the electrons. We start with the assumption that at the resonance peak at $7.3$\,T, the capture energy given by the blue line  is at the centre of the first Landau level for $B=7.3$\,T(Fig.\ref{figLand}a $\blacksquare$). It is also below the Fermi level given in red, which is essential, as there are no electrons for pump capture at energies above the source lead Fermi level.

\new{For our simulation we first model the scenario with no $E_c$ and  $B$ field dependency. The 0-DIP model suggests that $E_c$ is close to the inflection point energy $\alpha \Eout$, which is defined purely by the voltages applied to the gates of the QD as discussed earlier and is the leading term in $E_c$.} \new{Here} the capture energy of the electrons during the pump cycle \old{does not change as a function of magnetic field and we} is pinned \old{the capture energy}  at the centre of the lowest Landau level \old{$E_c = 1.51\times 10^{-21}$\,J} \new{$E_c = 9.4$\,meV } for all B  as shown in Fig. \ref{figLand}a by the dotted turquoise line. We then monitored the change in  the density of states at this fixed capture energy as a function of B, which is plotted as the dotted turquoise line in Fig.\ref{figLand}b. 

\old{Second,} \new{In an attempt to improve our simulation result we investigated  a linear dependency between $E_c$ and $B$. Here} we assumed that \new{ the magnetic dynamics in the quantum dot lead to a change in the energy of formation of the first confined state beyond 
 the inflection point, and to the first order, this change can be considered linear with} \old{the capture energy is changing as a function of} magnetic field.  We scaled the capture energy at $B = 7.2$\,T \old{($E_c = 1.51\times 10^{-21}$\, J)} \new{(\new{$E_c = 9.42$\,meV})} over the full $B$ field from $1$\,T to $9$\,T. The capture energy therefore changed by \old{$\Delta E_c = 3.9\times 10^{-22}$\,J} \new{$\Delta E_c = 2.4$\,meV} over the full range of B field. The plot of the density of states as the capture energy was changed  from  \old{$E_c = 1.21\times 10^{-21}$\,J} \new{$E_c = 7.5$\,meV} to \old{$ E_c=1.60\times 10^{-21}$\,J} \new{$ E_c=9.9$\,meV}  in step with the change in  B  from $1$\,T to $9$\,T, in multiples of $\Delta B = 9.0\times10^{-4}$\, T is shown by the dark blue solid trace in Fig.\ref{figLand}b.

From these two plots \new{in Fig.\ref{figLand}b} we see a much stronger coincidence  between the dark blue and red onset pumping curve allowing us to  conclude  \new{that} the capture energy of the electrons changes \new{linearly with B field, to a first approximation}, albeit slightly, over the full magnetic field sweep. \new{Further work will investigate high order dependencies between $E_c$ and $B$.\\
Due to the applied $B$ field there is a magnetic moment associated with the spin which can align parallel or anti-parallel with the field resulting in the lifting of the spin degeneracy. In our simulation this splitting in energy can be accounted for in the Landau level density of states calculation. In Fig.\ref{figLand}b, the blue and turquoise plots shown have taken spin into account. The inset given shows a comparison when spin is not included (pink curve) with the inclusion of spin (blue curve). Both  curves were produced for a linear dependency of $E_c$ with the $B$ field. There is a clear difference in the two curves over the field range. From the plot it is easy to conclude that the simulation without spin would not align to the pumping plateau curve in red, indicating that the inclusion of spin in the simulation is essential. It should be noted that there are states within the dot that would be magnetically confined\cite{}. These states are normally considered in static quantum dots where edge states in the leads are either reflected by the gates defining the dot or are transmitted through the dot allowing for electrons to enter. In our dynamic dot, as in the static case, electrons in lower Landau levels but with higher guiding centre potentials could be allowed into the dot versus electrons partitioned into higher Landau levels with a lower guiding potential. In our simulation we only studied the first pumped plateau corresponding to one electron per cycle. Future work will explore the confined magnetic states in the QD by focusing on plateaus with more than one electron.}


\section{Conclusion}
Our experimental data of single-electron pumping through a split-gate finger-gate quantum dot at high magnetic fields demonstrates oscillatory behaviour reminiscent of the SdH effect. The inclusion of a split-gate on the exit side of the quantum dot allows for an increase in gate resolution due to a lower lever-arm term\cite{Howe2021}. This, with the change in potential profile on the exit side of the quantum dot from a simple Gaussian to a saddle-point potential, has allowed for further probing of the dynamics of these on-demand electron sources. Our analysis together with a new \new{physical} \old{pumping} model \new{for the pumping dynamics} based on the foundations of the UDC model has allowed for the first reported direct measurement of the capture energy, the broadening of Landau levels in such pump devices and therefore the quantum lifetime of the pumped electrons. This demonstration of single electron pumping in the quantum Hall regime, where electrons are captured from specific Landau levels, shows the capture energy to be independent of the exit gate DC voltage for fixed B field, but a change in the capture energy is observed as the B field itself is varied.

Future work will explore further the frequency and temperature dependence of pumps operating in this regime as well the ejection energy of the pumped electrons\new{, where our simple model can serve as foundation for more involved theoretical work}. 
\section*{Acknowledgements}
This work was supported by the UK Engineering Physical Sciences Research Council, grants  EP/K040359/1 and EP/R029075/1, the National Research Foundation of South Africa. J.A.M. acknowledges a UKRI Future Leaders Fellowship, grant no. MR/X023125/1.

\section*{Author contributions}
E.P developed the theoretical model, produced data plots,  interpreted the data and contributed to the drafting of the paper. M.D.B ran the experiment, carried out the simulation work, interpreted the data and contributed to the drafting of the paper. Both E.P and M.D.B contributed equally.  H.H. designed, processed the devices and assisted with measurements. J.A.M contributed to data interpretation.  T.M carried out the e-beam lithography. H.E.B and D.A.R developed and grew the GaAs/AlGaAs heterostructures used. M.P. provided support, references, and useful discussions. All authors reviewed the manuscript. 

\section*{Data availability}
All experimental data and detailed simulation results presented in the study are available from the corresponding author upon reasonable request.

\section*{Code availability}
The code employed to generate the simulations presented in this work, as well as the data analysis code are available upon
request. Interested parties may inquire about access to the code by contacting the corresponding author.

\section*{Competing interests}
The authors declare no competing interests.

\section*{Methods}
\subsection{Fabrication and Experiment Setup}
\label{SI-fab}
Our device comprises a single-electron pump with two Ti/Au gates in a finger-gate split-gate geometry (red and green in Fig.\ref{fig1}a respectively). Only the entrance gate, in our the case the finger-gate, is coupled directly to the Agilent RF signal generator. The pumped current is measured via a Keithley 6430 Source-Measure unit capable of measuring at the femto-ampere level. 
DC voltages are applied to the gates via multiple channel 16 bit DAQs. Unlike the earlier pump configurations, in which oscillating voltages have been applied to both the exit and entrance gate\cite{Fujiwara2004, Blumenthal2007}, we follow a later protocol\cite{Fujiwara2008, Kaestner2008}, in which the potential of the exit gate is held constant during the pump cycle, and the entrance gate oscillates due to the RF signal added to the constant potential via a bias T.
The gates were fabricated using electron beam lithography with Ti/Au thermally evaporated. The entrance finger gate (red) and split gate (green) are used to define the QD entrance and exit gate respectively. DC voltages were applied to all gates using a NI2969 cDAQ with the RF signal coupled to the exit gate via a bias tee. The RF source used was a HP E4400B driving a simple sinusoidal wave. 

The device was fabricated on a two-dimensional high mobility GaAs/Al\textsubscript{x}Ga\textsubscript{1-x}As Si-doped electron gas system grown using molecular beam epitaxy. The 2DEG was formed 90\,nm below the surface (10\,nm GaAs cap, 40\,nm Si-doped GaAs/Al\textsubscript{x}Ga\textsubscript{1-x}As, 40 \,nm GaAs/Al\textsubscript{x}Ga\textsubscript{1-x}As spacer, and GaAs substrate with a carrier density $n = 1.9\times 10^{11}$\,cm\textsuperscript{-2} and mobility $\mu = 1.014\times 10^6$\,cm\textsuperscript{2}/Vs). The 2DEG channel pattern was defined using electron-beam lithography (EBL) and etched to a depth of 40\,nm using wet chemistry. The split gate has a width and gap of $400$\,nm, whilst the finger gate a width of $150$\,nm. The pitch of the gates are $200$\,nm. \new{The sample contained four AuGeNi ohmic contacts. Two were utilised for measurement of pumped current and four when measuring the SdH effect}
The sample was loaded into a Leiden Cryogenics dilution fridge with a base temperature of $7$\,mK and a $10$\,T superconducting magnet.  The pumped current was measured using a Keithley 6430 source measure unit (SMU), which was connected to the drain side of the pump. The source side was grounded.

\new{The SdH plot without the RF signal (Fig.\ref{fig1}, turquoise) differs from assessment measurements in the same wafer with a standard Hall bar. This difference can be attributed to the geometry of the mesa in the pump deviating substantially from a standard Hall bar geometry. The inclusion of Ti/Au gates as well as the fabrication process itself also affect the wafer, altering the electron density and mobility. The application of the RF signal further influences the magnetic field dependence of the device resistance, however both retain the general peak structure of the signature SdH.}

\subsection{Dot parameters -- analytical consideration}
We find the dependence of the quantum dot parameters $W_d$ and $D_d$ -- width of the potential barrier and the dot depth in a 1D case -- shortly after the system has passed the 0-DIP configuration. The potential profile $E(x)$ is the sum of the potential profiles of both electrodes:
\begin{equation}
    E(x)=\Ein \fin +\Eout \fout
\end{equation}
where $\fin$ and $\fout$ are the normalised functions of the potentials creates by each of the electrodes independently, we assume they are smooth, differentiable, with a simple peak, and $\max (\fin) = \max (\fout) = 1$.

In the 0-DIP configuration, $\Ein=\alpha \Eout$. If $x_0$ is the inflection point, then the first and second derivatives of the potential profile in it are equal to zero:
\begin{equation}
    \begin{dcases}
        E'(x_0)=\alpha \Eout \fin'(x_0)+\Eout \fout'(x_0)=0 
        \\[5pt]
        E''(x_0)=\alpha \Eout \fin''(x_0)+\Eout \fout''(x_0)=0
    \end{dcases}
\end{equation}
 Shortly past the 0-DIP configuration, $\Ein=\alpha \Eout + \dd E$, the derivatives at the same point are:
\begin{equation}
    \begin{dcases}
        E'(x_0)= \dd E \fin'(x_0) 
        \\[5pt]
        E''(x_0)= \dd E \fin''(x_0)
    \end{dcases}
\end{equation}
In this case (see Fig.\ref{fig3}d), the potential has three special points -- the new inflection point $x_i$, the local maximum $x_+$ and the local minimum $x_-$. Assuming that $\dd E$ is small and the special points lie within a small vicinity of the former inflection point $x_0$ we find the distance between the new and the former inflection points $\dd x_i$ in the Taylor expansion. Postulating $E''(x_0+\dd x_i)=0$, we arrive at:
\begin{equation}
    \dd x_i= - \frac{\dd E \fin''(x_0)}{\alpha \Eout \fin'''(x_0)+\Eout \fout'''(x_0)}
\end{equation}
At the new inflection point, to the first order,
\begin{equation}
    \begin{dcases}
        E'(x_i)=\dd E \fin'(x_0)
        \\[5pt]
        E''(x_i)=0
        \\[5pt]
        E'''(x_i)=\alpha \Eout \fin'''(x_0)+\Eout \fout'''(x_0)
    \end{dcases}
\end{equation}
Finally, we find the distance between the new inflection point $x_i$ and the maximum and minimum $x_+$ and $x_-$. Setting
\begin{equation}
    E'(x_i+\dd x)=E'(x_i)+ E''(x_i) \dd x +E'''(x_i) \frac{\dd x^2}{2}=0
\end{equation}
we arrive at:
\begin{equation}
 \dd x = \sqrt{- \frac{2 \dd E \fin'(x_0)}{\alpha \Eout \fin'''(x_0)+\Eout \fout'''(x_0)}}
\end{equation}
We set $2 \dd x$, the distance between the local minimum and maximum as the characteristic size of the dot, which is equal to the width of the potential barrier; and the depth of the dot, or the height of the barrier can be found as:
\begin{equation}
    D_d=2E'(x_i)\dd x + 2 E'''(x_0) \frac{\dd x^3}{6}
\end{equation}
which yields:
\begin{equation}
    D_d=\frac{4}{3}\dd E \fin'(x_0)\sqrt{- \frac{2 \dd E \fin'(x_0)}{\alpha \Eout \fin'''(x_0)+\Eout \fout'''(x_0)}}
\end{equation}
The final dependence on the external parameters has the form:
\begin{equation}
    \begin{dcases}
    W_d=C_W \sqrt{\frac{\dd E}{\Eout}}\\[5pt]
    D_d=C_D \sqrt{\frac{(\dd E)^3}{\Eout}}
    \end{dcases}
\end{equation}.

%

\end{document}